\documentclass[aps,prl,twocolumn]{revtex4}
\usepackage{amsfonts}
\usepackage{amsmath}
\usepackage{color}
\usepackage[pdftex,colorlinks=false,bookmarks=false]{hyperref}
\usepackage{amssymb}
\usepackage{graphicx}

\begin{document}

\title{Scalable Quantum Networks based on Few-Qubit Registers}
\author{Liang Jiang$^{1}$, Jacob M. Taylor$^{1,2}$, Anders S. S\o rensen$^{3}
$, Mikhail D. Lukin$^{1}$}
\affiliation{$^{1}$ Department of Physics, Harvard University, Cambridge, Massachusetts
02138}
\affiliation{$^{2}$ Department of Physics, Massachusetts Institute of Technology,
Cambridge, Massachusetts 02139}
\affiliation{$^{3}$ Quantop and The Niels Bohr Institute, University of Copenhagen,
DK-2100 Copenhagen \O , Denmark}
\date{\today}

\begin{abstract}
We describe and analyze a hybrid approach to scalable quantum computation
based on an optically connected network of few-qubit quantum registers. We
show that probabilistically connected five-qubit quantum registers suffice
for deterministic, fault-tolerant quantum computation even when state
preparation, measurement, and entanglement generation all have substantial
errors. We discuss requirements for achieving fault-tolerant operation for
two specific implementations of our approach.
\end{abstract}

\maketitle

The key challenge in experimental quantum information science is to identify
isolated quantum mechanical systems with good coherence properties that can
be manipulated and coupled together in a scalable fashion. Substantial
progress has been made towards the physical implementation of few-qubit
quantum registers using systems of coupled trapped ions~\cite%
{Cirac04,Leibfried03, Haffner05, Reichle06}, superconducting islands \cite%
{Yamamoto03,Wallraff04}, solid-state qubits based on electronic spins in
semiconductors \cite{Petta05}, and color centers in diamond~\cite%
{Wrachtrup01,Jelezko04,Childress06,Dutt06}. While the precise manipulation
of large, multi-qubit systems still remains an outstanding challenge,
approaches for connecting such few qubit registers into large scale circuits
are currently being explored both theoretically \cite%
{Cirac97,Dur03,Lim06,Oi06,vanMeter06,Duan04} and experimentally \cite%
{Legero04,Birnbaum05}. Of specific importance are approaches which can yield
fault-tolerant operations with minimal resources and realistic (high) error
rates.

\begin{figure}[b]
\begin{center}
\includegraphics[
width=7cm
]{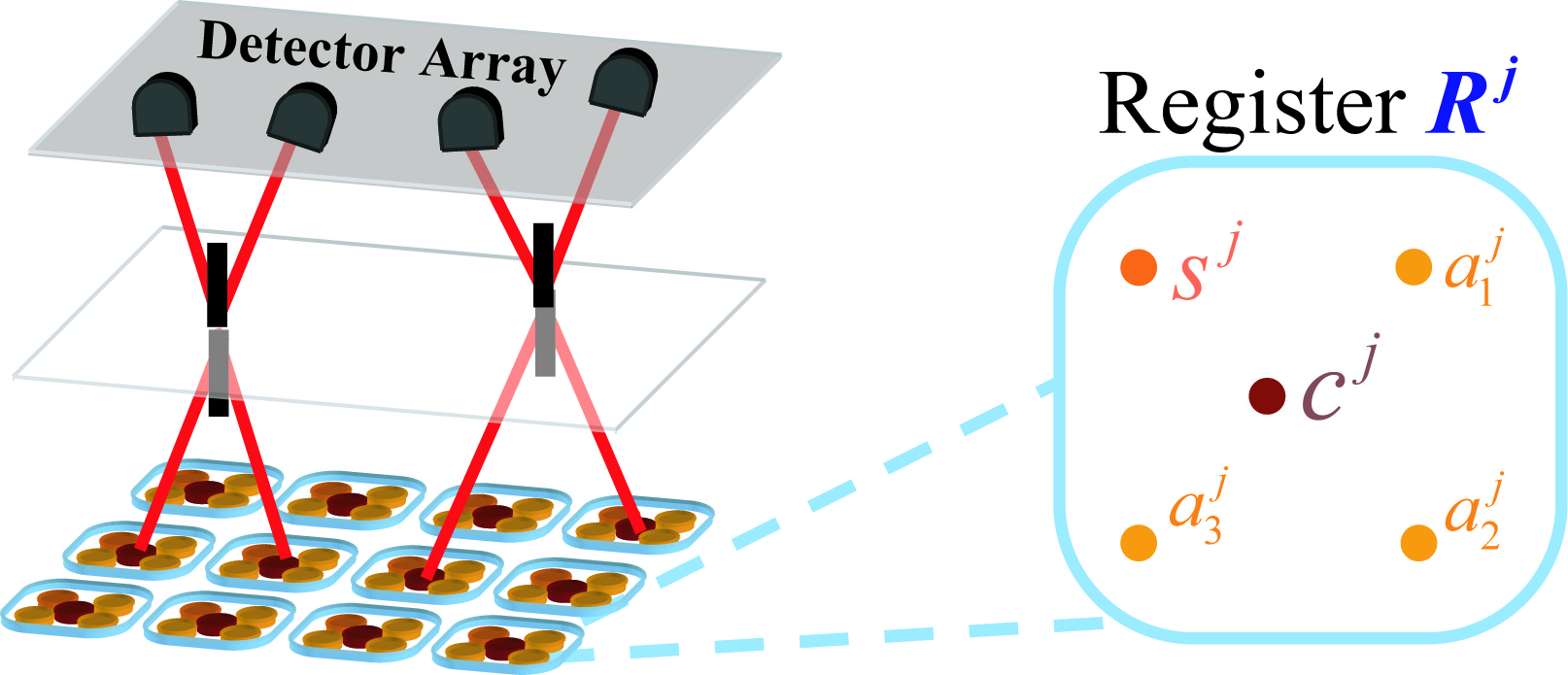}
\end{center}
\caption[fig1]{(color online). Illustration of distributed quantum computer
based on many quantum registers. Each register has five physical qubits,
including one communication qubit ($c$), one storage qubit ($s$), and three
auxiliary qubits ($a_{1,2,3}$). Local operations for qubits from the same
register have high fidelity. Entanglement between remote registers can be
generated probabilistically \protect\cite{Childress05,Duan03}. Optical MEMS
devices \protect\cite{Kim07} can efficiently route photons and couple
arbitrary pair of registers. Detector array can \emph{simultaneously}
generate entanglement for many pairs of registers. }
\label{f:distributedQC}
\end{figure}

In Ref.~\cite{Dur03} a novel technique to scalable quantum computation was
suggested, where high fidelity local operations can be used to correct low
fidelity non-local operations, using techniques that are currently being
explored for quantum communication \cite{Briegel98,Dur99,Childress05}. In
this Letter, we present a hybrid approach, which requires only $5$ (or
fewer)-qubit registers with local determinstic coupling, while providing
additional improvements over the earlier protocol \cite{Dur03}: reduced
measurement errors, higher fidelity, and more efficient entanglement
purification. The small registers are connected by optical photons, which
enables non-local coupling gates and reduces the requirement for fault
tolerant quantum computation \cite{Svore05}. Specifically, we analyze two
physical systems where this approach is very effective. We consider an
architecture where pairwise non-local entanglement can be created in
parallel, as indicated in Fig.~\ref{f:distributedQC}. This is achieved via
simultaneous optical excitation of the selected register pairs followed by
photon-detection in specific channel. We use a Markov chain analysis to
estimate the overhead in time and operational errors, and discuss the
feasibility of large scale, fault-tolerant quantum computation using this
approach.

The present work is motivated by experimental advances in two specific
physical systems. Recent experiments have demonstrated quantum registers
composed of few trapped ions, which can support high-fidelity local
operations \cite{Leibfried03, Haffner05, Reichle06}. The ion qubits can
couple to light efficiently \cite{Blinov04} and were recognized early for
their potential in an optically coupled component~\cite{Dur03,Duan04}.
Probabilistic entanglement of remote ion qubits mediated by photons has also
been demonstrated \cite{Moehring07,Maunz07}. At the same time, few-qubit
quantum registers have been recently implemented in high-purity diamond
samples \cite{Jelezko04,Childress06,Dutt06}. Here, quantum bits are encoded
in individual nuclear spins, which are extraordinarily good quantum memories
\cite{Dutt06} and can also be manipulated with high precision using
techniques from NMR \cite{Vandersypen04}. The electronic spin associated
with a nitrogen-vacancy (NV) color center enables addressing and
polarization of nuclei, and entanglement generation between remote
registers. While for systems of trapped ions there exist several approaches
for coupling remote few-qubit registers (such as those based on moving the
ions \cite{Kielpinski02}), for NV centers in diamond it is difficult to
conceive a direct construction of large scale multi-qubit systems without
major advances in fabrication technology. For the latter scenario the hybrid
approach developed here is required. Furthermore the use of light has the
major advantage that it allows for connecting qubits over long distances,
which reduces the requirement for fault-tolerant quantum computation \cite%
{Svore05}.

%computing may
%be ideal. Furthermore the use of light to implement gates allows gates over
%long-distances, which can reduce the threshold for fault-tolerant quantum computation.

%For the general case, even both entanglement generation and measurement may
%have low fidelity (say 0.95), it is still possible to achieve fault-tolerant
%quantum computation, as long as local unitary operations are good ($p_{L}%
%\leq10^{-6}$ error probabilities).

%For the general case, we find that with good local operations ($p_{L}
%\leq10^{-6}$ error probabilities) and low fidelity entanglement generation ($F
%= 0.9$), large scale, fault tolerant computations may be possible using
%quantum error correcting codes with 50 physical qubits (10 registers) per
%logical qubit and a factor of 1000 overhead in logical gate time to the
%entanglement generation time.

%\paragraph{Quantum register.}

We define a \emph{quantum register} as a few-qubit device that contains one
\emph{communication} qubit, with a photonic interface; one \emph{storage}
qubit, with very good coherence times; and several \emph{auxiliary} qubits,
used for purification and error correction (described below).
%A critical requirement for a quantum register is high-fidelity unitary operations between the qubits within a register.

The simplest quantum register requires only two qubits: one for storage and
the other for communication. Entanglement between two remote registers may
be generated using probabilistic approaches from quantum communication (\cite%
{Childress05} and references therein). In general, such entanglement
generation produces a Bell state of the communication qubits from different
registers, conditioned on certain measurement outcomes. If state generation
fails, it can be re-attempted until success, with an exponentially
decreasing chance of continued failure. When the communication qubits ($c^{1}
$ and $c^{2}$) are prepared in the Bell state, we can immediately perform
the remote C-NOT gate on the storage qubits ($s^{1}$ and $s^{2}$) using the
gate-teleportation circuit between registers $R^{1}$ and $R^{2}$. This can
be accomplished \cite{Gottesman99,Eisert00,Duan04} via a sequence of local
C-NOTs within each register, followed by measurement of two communication
qubits and subsequent local rotations. Since arbitrary rotations on a single
qubit can be performed within a register, the C-NOT operation between
different quantum registers is in principle sufficient for universal quantum
computation. Similar approaches are also known for deterministic generation
of graph states \cite{Benjamin06} ---an essential resource for one-way
quantum computation \cite{Raussendorf00}.

\begin{figure}[ptb]
\begin{center}
\includegraphics[
width=8.0cm
]{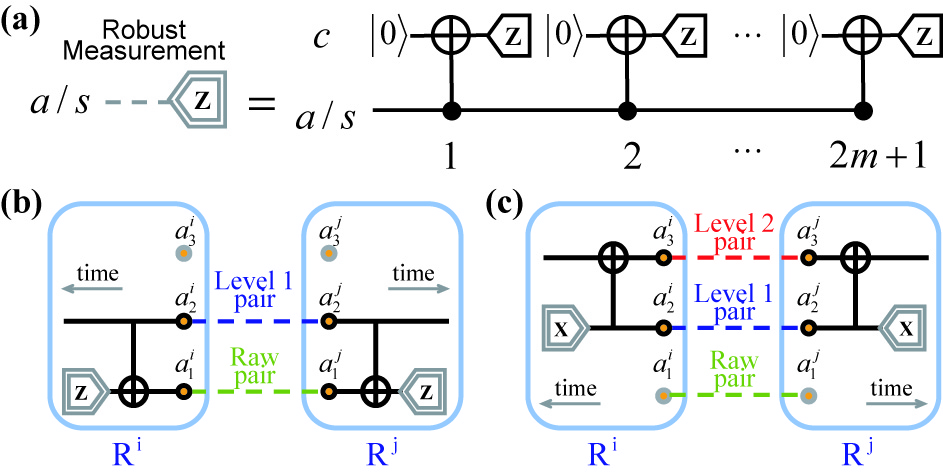}
\end{center}
\caption[fig2]{(color online). Circuits for robust operations. (a) Robust
measurement of the auxiliary/storage qubit, $a/s$, based on majority vote
from $2m+1$ outcomes of the communication qubit, $c$. Robust measurement is
denoted by the box shown in the upper left corner. (b)(c) Using entanglement
pumping to create high fidelity entangled pairs between two registers $R^{i}$
and $R^{j}$. If the two outcomes are the same, it is a successful step of
pumping; otherwise generate new pairs and restart the pumping operation from
the beginning. The two circuits are for the first level pumping and the
second level pumping, purifying bit- and phase-errors, respectively.}
\label{f:circuits}
\end{figure}
%EndExpansion

In practice, the qubit measurement, initialization, and entanglement
generation can be fairly noisy with error probabilities as high as a few
percent, due to practical limitations such as finite collection efficiency
and poor interferometric stability. As a result the corresponding error
probability in non-local gate circuit will also be very high. In contrast,
local unitary operations may fail infrequently ($p_{L}\lesssim10^{-4}$) when
quantum control techniques for small quantum system are utilized \cite%
{Vandersypen04,Leibfried03}. We now show that the most important sources of
imperfections, such as imperfect initialization, measurement errors for
individual qubits in each quantum register, and entanglement generation
errors between registers, can be corrected with a modest increase in
register size. We determine that with just \emph{three} additional auxiliary
qubits and high-fidelity local unitary operations, all these errors can be
efficiently suppressed by bit-verification and entanglement purification
\cite{Briegel98,Dur99}. This provides an extension of Ref.~\cite{Dur03} that
mostly focused on suppressing errors from entanglement generation.

We are assuming in the following a separation of error probabilities: any
internal, unitary operation of the register fails with low probability, $%
p_{L}$, while all operations connecting the communication qubit to the
outside world (initialization, measurement, and entanglement generation)
fail with error probabilities that can be several orders of magnitude
higher. For specificity, we set these error probabilities to $p_{I}$, $p_{M}$%
, and $1-F$, respectively. In terms of these quantities the error
probability in the non-local C-NOT gate circuit is of order $%
p_{CNOT}\sim(1-F)+2p_{L}+2p_{M}$. We now show how this fidelity can be
greatly increased.

\emph{Robust measurement} can be implemented by bit-verification: a majority
vote among the measurement outcomes (Fig.~\ref{f:circuits}a), following a
sequence of C-NOT operations between the auxiliary/storage qubit and the
communication qubit. This also allows \emph{robust initialization} by
measurement. High-fidelity \emph{robust entanglement generation} is achieved
via entanglement purification \cite{Briegel98,Dur99,Dur03} (Fig.~\ref%
{f:circuits}bc), in which lower fidelity entanglement between the
communication qubits is used to purify entanglement between the auxiliary
qubits, which can then be used for the remote C-NOT operation. To make the
most efficient use of physical qubits, we introduce a new two-level
entanglement pumping scheme. Our circuit (Fig.~\ref{f:circuits}b) uses raw
Bell pairs to repeatedly purify (\textquotedblleft pump\textquotedblright)
against bit-errors, then the bit-purified Bell pairs are used to pump
against phase-errors (Fig.~\ref{f:circuits}c).

Entanglement pumping, like entanglement generation, is probabilistic;
however, failures are detected. Still, in computation, where each logical
gate should be completed within the allocated time (clock cycle), failed
entanglement pumping can lead to gate failure. To demonstrate the
feasibility of our approach for quantum computation, we next analyze the
time required for robust initialization, measurement and entanglement
generation, and show that the failure probability for these procedures can
be made sufficiently small with reasonable time overhead.

%\paragraph{Robust initialization}

The measurement circuit shown in Fig.~\ref{f:circuits}a yields the correct
result based on majority vote from $2m+1$ consecutive readouts
(bit-verification). Since the evolution of the system (C-NOT gate) commutes
with the measured observable ($Z$ operator) of the auxiliary/storage qubit,
it is a quantum non-demolition (QND) measurement, which can be repeated many
times. The error probability for majority vote measurement scheme is:%
\begin{equation}
\varepsilon_{M}\approx\left(
\begin{array}{c}
2m+1 \\
m+1%
\end{array}
\right) \left( p_{I}+p_{M}\right) ^{m+1}+\frac{2m+1}{2}p_{L}.   \label{em}
\end{equation}
Suppose $p_{I}=p_{M}=5\%$, we can achieve $\varepsilon_{M}\approx
8\times10^{-4}$ by choosing $m^{\ast}=6$ for $p_{L}=10^{-4}$, or even $%
\varepsilon_{M}\approx12\times10^{-6}$ for $m^{\ast}=10$ and $p_{L}=10^{-6}$%
. Recently, measurement with very high fidelity ($\varepsilon_{M}$ as low as
$6\times10^{-4}$) has been demonstrated in the ion-trap system \cite{Hume07}%
, using similar ideas as above.
%\emph{ For convenience of discussion, we shall add }$\tilde{\varepsilon}_{M}$\emph{ to
%the set of imperfection parameters--- }$\left(  1-F,p_{I},p_{M},p_{L}%
%,\tilde{\varepsilon}_{M}\right)  $\emph{.}
The time for robust measurement is%
\begin{equation}
\tilde{t}_{M}=\left( 2m+1\right) \left( t_{I}+t_{L}+t_{M}\right) ,
\label{tm}
\end{equation}
where $t_{I}$, $t_{L}$, and $t_{M}$ are times for initialization, local
unitary gate, and measurement, respectively.

%\paragraph{Robust entanglement generation}

We now use robust measurement and entanglement generation to perform
entanglement pumping. Suppose the raw Bell pairs have initial fidelity $%
F=\left\langle \left\vert \Phi^{+}\right\rangle \!\left\langle \Phi
^{+}\right\vert \right\rangle $ due to \emph{depolarizing error}. We apply
two-level entanglement pumping. The first level has $n_{b}$ steps of
bit-error pumping using raw Bell pairs (Fig.~\ref{f:circuits}b) to produce a
bit-error-purified entangled pair. The second level uses these
bit-error-purified pairs for $n_{p}$ steps of phase-error pumping (Fig.~\ref%
{f:circuits}c).

For successful purification, the infidelity of the purified pair, $%
\varepsilon_{E,\text{\textrm{infid}}}^{\left( n_{b},n_{p}\right) }$, depends
on both the control parameters $\left( n_{b},n_{p}\right) $ and the
imperfection parameters $\left( F,p_{L},\varepsilon_{M}\right) $. For
depolarizing error, we find%
\begin{align*}
& \varepsilon_{E,\text{\textrm{infid}}}^{\left( n_{b},n_{p}\right) }\approx%
\frac{3+2n_{p}}{4}p_{L}+\frac{4+2\left( n_{b}+n_{p}\right) }{3}\left(
1-F\right) \varepsilon_{M} \\
& +\left( n_{p}+1\right) \left( \frac{2\left( 1-F\right) }{3}\right)
^{n_{b}+1}+\left( \frac{\left( n_{b}+1\right) \left( 1-F\right) }{3}\right)
^{n_{p}+1}
\end{align*}
to the leading order of $p_{L}$ and $\varepsilon_{M}$. The dependence on the
initial infidelity $1-F$ is exponentially suppressed at the cost of a linear
increase of error from local operations $p_{L}$ and robust measurement $%
\varepsilon_{M}$. Measurement-related errors are suppressed by the prefactor
$1-F$, since measurement error does not cause infidelity unless combined
with other errors. In the limit of ideal operations ($p_{L},\varepsilon
_{M}\rightarrow0$), the infidelity $\varepsilon_{E,\text{\textrm{infid}}%
}^{\left( n_{b},n_{p}\right) }$ can be arbitrarily close to zero \cite%
{JTSL07b}. On the other hand, if we use the standard entanglement pumping
scheme \cite{Briegel98,Dur99} (that alternates purification of bit and phase
errors within each pumping level), the reduced infidelity from two-level
pumping is always larger than $\left( 1-F\right) ^{2}/9$. Therefore, for
very small $p_{L}$ and $\varepsilon_{M}$, the new pumping scheme is crucial
to minimize the number of qubits per register.

The overall success probability can be defined as the joint probability that
all successive steps succeed. We use the model of finite-state Markov chain
\cite{Meyn93} to directly calculate the \emph{failure probability} of $%
\left( n_{b},n_{p}\right) $-two-level entanglement pumping using $N_{\mathrm{%
tot}}$ raw Bell pairs, denoted as $\varepsilon_{E,fail}^{\left(
n_{b},n_{p}\right) }\left( N_{\mathrm{tot}}\right) $. See Ref.~\cite{JTSL07b}
for detailed analysis.

For given $F$, $p_{L}$, and $\varepsilon_{M}$, the purified pair has minimum
infidelity $\Delta_{\min}=\varepsilon_{E,\text{\textrm{infid}}}^{\left(
n_{b}^{\ast},n_{p}^{\ast}\right) }$, obtained by the optimal choice of the
control parameters $\left( n_{b}^{\ast},n_{p}^{\ast}\right) $. Then, we
calculate the typical value for $N_{\mathrm{tot}}$, by requiring the failure
probability and the minimum infidelity to be equal, $\varepsilon _{E,\text{%
\textrm{fail}}}^{\left( n_{b}^{\ast},n_{p}^{\ast}\right) }\left( N_{\mathrm{%
tot}}\right) =\Delta_{\min}$. The total error probability is
\begin{equation}
\varepsilon_{E}\approx\varepsilon_{E,\text{\textrm{fail}}}^{\left(
n_{b}^{\ast},n_{p}^{\ast}\right) }\left( N_{\mathrm{tot}}\right)
+\Delta_{\min}=2\Delta_{\min}.
\end{equation}
We remark that a faster and less resource intensive approach may be used if
the unpurified Bell pair is dominated by \emph{dephasing error}. And
one-level pumping may be sufficient (i.e. no bit-error purification, $n_{b}=0
$ %\footnote[15]{For dephasing error, we have $\varepsilon_{E}^{\left(
%0,n_{p}\right)  }\approx\left(  1-F\right)  ^{n_{p}+1}+\frac{2+n_{p}}{4}%
%p_{L}+2\left(  1-F\right)  \varepsilon_{M}$ by expanding to the leading order
%of $p_{L}$ and $\varepsilon_{M}$}
). The total time for robust entanglement generation $\tilde{t}_{E}$ is
\begin{equation}
\tilde{t}_{E}\approx\left\langle N_{\mathrm{tot}}\right\rangle \times\left(
t_{E}+t_{L}+\tilde{t}_{M}\right) ,
\end{equation}
where $t_{E}$ is the average generation time of the unpurified Bell pair.

\begin{figure}[ptb]
\begin{center}
\includegraphics[
width=8.5  cm
]{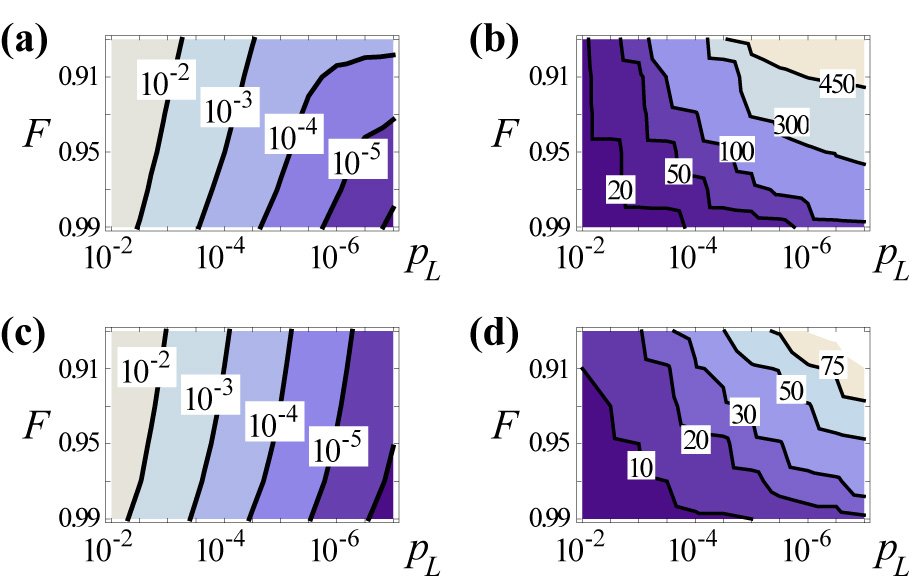}
\end{center}
\caption[fig3]{(color online). Contours of the total error probability after
purification $\protect\varepsilon_{E}$ (a,c) and total number of unpurified
Bell pairs $N_{\mathrm{tot}}$ (b,d) with respect to the imperfection
parameters $p_{L}$ (horizontal axis) and $F$ (vertical axis). (a,b)
Two-level pumping is used for depolarizing error, and (c,d) one-level
pumping for dephasing error. $p_{I}=p_{M}=5\%$ is assumed. }
\label{f:contours}
\end{figure}

Figure \ref{f:contours} shows the contours of $\varepsilon_{E}$ and $N_{%
\mathrm{tot}}$ with respect to the imperfection parameters $p_{L}$ and $1-F$%
. We assume $p_{I}=p_{M}=5\%$ for the plot. The choice of $p_{I}$ and $p_{M}$
($<10\%$) has marginal effect to the contours, since they only modifies $%
\varepsilon_{M}$ marginally. For initial fidelity $F_{0}>0.95$, the contours
of $\varepsilon_{E}$ are almost vertical; that is $\varepsilon_{E}$ is
mostly limited by $p_{L}$ with an overhead factor of about $10$. The
contours of $N_{tot}$ indicate that the entanglement pumping needs about
tens or hundreds of raw Bell pairs to ensure a very high success probability.

We introduce the \emph{clock cycle time} $t_{C}=\tilde{t}_{E}+2t_{L}+\tilde {%
t}_{M}\approx\tilde{t}_{E}$\ and the \emph{effective error probability} $%
\gamma=\varepsilon_{E}+2p_{L}+2\varepsilon_{M}$ for general coupling gate
between two registers, which can be implemented with a similar approach as
the remote C-NOT gate \cite{Eisert00}. We now provide an estimate of clock
cycle time based on realistic parameters. The time for optical
initialization/measurement is $t_{I}=t_{M}\approx\frac{\ln p_{M}}{\ln\left(
1-\eta\right) }\frac{\tau}{C}$, with photon collection/detection efficiency $%
\eta$, vacuum radiative lifetime $\tau$, and the Purcell factor $C$ for
cavity-enhanced radiative decay. We assume that entanglement is generated
based on detection of two photons \cite{Duan03}, which takes time $%
t_{E}\approx\left( t_{I}+\tau/C\right) /\eta^{2}$. If the bit-errors are
efficiently suppressed by the intrinsic purification of the entanglement
generation scheme, one-level pumping is sufficient; otherwise two-level
pumping is needed. Suppose the parameters are $\left( t_{L},\tau
,\eta,C\right) =\left( 0.1\text{ }\mu\text{s, }10\text{ ns, }0.2\text{, }%
10\right) $ \cite{Garcia03,Steinmetz06,Keller04} and $\left(
1-F,p_{I},p_{M},p_{L},\varepsilon_{M}\right) =\left(
5\%,5\%,5\%,10^{-6},12\times10^{-6}\right) $. For depolarizing errors,
two-level pumping can achieve $\left( t_{C}\text{,}\gamma\right) =\left( 997%
\text{ }\mu\text{s, }4.5\times10^{-5}\right) $. If all bit-errors are
suppressed by the intrinsic purification of the coincidence scheme,
one-level pumping is sufficient and $\left( t_{C}\text{,}\gamma\right)
=\left( 140\text{ }\mu\text{s, }3.4\times10^{-5}\right) $. Finally, $t_{C}$
should be much shorter than the memory time of the storage qubit, $t_{mem}$.
This is indeed the case for both trapped ions (where $t_{mem}\sim10$ s has
been demonstrated \cite{Langer05,Haffner05b}) and proximal nuclear spins of
NV\ centers (where $t_{mem}$ approaches $1$ s \cite{Dutt06}) \cite{JTSL07b}.

This approach yields gates between quantum registers %which, combined with
%local unitary operations (error probability $p_{L}$) and robust measurement
%($\tilde{\epsilon}_{M}$), allow
to implement arbitrary quantum circuits. Errors can be further suppressed by
using quantum error correction. For example, as shown in Fig.~\ref%
{f:contours}, $\left( p_{L},F\right) =\left( 10^{-4},0.95\right) $ can yield
$\gamma\leq2\times10^{-3}$, well below the $1\%$ threshold for fault
tolerant computation for approaches such as the $C_{4}/C_{6}$ code~\cite%
{Knill05} or 2D toric codes~\cite{Raussendorf07}; $\left( p_{L},F\right)
=\left( 10^{-6},0.95\right) $ can achieve $\gamma\leq5\times10^{-5}$, which
allows efficient codes such as the BCH [[127,43,13]] code to be used without
concatenation. Following Ref.~\cite{Steane03}, we estimate $10$ registers
per logical qubit to be necessary for a calculation involving $10^{4}$
logical qubits and $10^{6}$ logical gates.

In conclusion, we have analyzed a hybrid approach to fault-tolerant quantum
computation with optically coupled few-qubit quantum registers.
%With a reasonable overhead in operational time and gate error probabilities,
%this approach enables the reduction of an experimental challenge of building a
%thousand-qubit quantum computer into a more feasible task of optically
%coupling five-qubit quantum registers.
We further note that it is possible to facilitate fault-tolerant quantum
computation with special operations from the hybrid approach such as partial
Bell measurement \cite{JTSL07b} or with systematic optimization using
dynamic programming \cite{Jiang07}.

The authors wish to thank Gurudev Dutt, Lily Childress, Paola Cappellaro,
Phillip Hemmer, and Charles Marcus. This work is supported by NSF, DTO,
ARO-MURI, the Packard Foundations, Pappalardo Fellowship, and the Danish
Natural Science Research Council.


\begin{thebibliography}{99}
\bibitem{Cirac04} J. I. Cirac and P. Zoller, Physics Today \textbf{57}, 38
(2004).

\bibitem{Leibfried03} D. Leibfried, \textit{et al.}, Nature (London) \textbf{%
422}, 412 (2003).

\bibitem{Haffner05} H. Haffner, \textit{et al.}, Nature (London) \textbf{438}%
, 643 (2005).

\bibitem{Reichle06} R. Reichle, \textit{et al.}, Nature (London) \textbf{443}%
, 838 (2006).

\bibitem{Yamamoto03} T. Yamamoto, \textit{et al.}, Nature (London) \textbf{%
421}, 343 (2003).

\bibitem{Wallraff04} A. Wallraff, \textit{et al.}, Nature (London) \textbf{%
431}, 162 (2004).

\bibitem{Petta05} J. R. Petta, \textit{et al.}, Science \textbf{309}, 2180
(2005).

\bibitem{Wrachtrup01} J. Wrachtrup, \textit{et al.}, Opt. Spectrosc. \textbf{%
91}, 429 (2001).

\bibitem{Jelezko04} F. Jelezko, \textit{et al.}, Phys. Rev. Lett. \textbf{93}%
, 130501 (2004).

\bibitem{Childress06} L. Childress, \textit{et al.}, Science \textbf{314},
281 (2006).

\bibitem{Dutt06} M. V. G. Dutt, \textit{et al.}, Science \textbf{316}, 1312
(2007).

\bibitem{Cirac97} J. I. Cirac, et al., Phys. Rev. Lett. \textbf{78}, 3221
(1997).

\bibitem{Dur03} W. D\"ur and H. J. Briegel, Phys. Rev. Lett. \textbf{90},
067901 (2003).

\bibitem{Duan04} L. M. Duan, \textit{et al.}, Quantum Inf. Comput. \textbf{4}%
, 165 (2004).

\bibitem{Lim06} Y. L. Lim, \textit{et al.}, Phys. Rev. A \textbf{73}, 012304
(2006).

\bibitem{Oi06} D. K. L. Oi, \textit{et al.}, Phys. Rev. A \textbf{74},
052313 (2006).

\bibitem{vanMeter06} R. van Meter, \textit{et al.}, quant-ph/0607160 (2006).

\bibitem{Legero04} T. Legero, \textit{et al.}, Phys. Rev. Lett. \textbf{93},
070503 (2004).

\bibitem{Birnbaum05} K. M. Birnbaum, \textit{et al.}, Nature (London)
\textbf{436}, 87 (2005).

\bibitem{Briegel98} H.-J. Briegel, \textit{et al.}, Phys. Rev. Lett. \textbf{%
81}, 5932 (1998).

\bibitem{Dur99} W. Dur, \textit{et al.}, Phys. Rev. A \textbf{59}, 169
(1999).

\bibitem{Childress05} L. Childress, \textit{et al.}, Phys. Rev. A \textbf{72}%
, 052330 (2005), Phys. Rev. Lett. \textbf{96}, 070504 (2006).

\bibitem{Svore05} K. M. Svore, \textit{et al.}, Phys. Rev. A 72, 022317
(2005).

\bibitem{Blinov04} B. B. Blinov, \textit{et al.}, Nature (London) \textbf{428%
}, 153 (2004).

\bibitem{Moehring07} D. L. Moehring, et al., Nature (London) 449, 68 (2007).

\bibitem {Maunz07}P. Maunz, \textit{et al.}, Nature Phys. \textbf{3}, 538
(2007).
%doi:10.1038/nphys644 (2007).

\bibitem{Vandersypen04} L. M. K. Vandersypen and I. L. Chuang, Rev. Mod.
Phys. \textbf{76}, 1037 (2004).

\bibitem{Kielpinski02} D. Kielpinski, \textit{et al.}, Nature (London)
\textbf{417}, 709 (2002).

\bibitem{Duan03} L. M. Duan and H. J. Kimble, Phys. Rev. Lett. \textbf{90},
253601 (2003). C. Simon and W. T. M. Irvine, Phys. Rev. Lett. \textbf{91},
110405 (2003).

\bibitem{Kim07} C. Kim, \textit{et al.}, Ieee Journal of Selected Topics in
Quantum Electronics 13, 322 (2007).

\bibitem{Gottesman99} D. Gottesman and I. L. Chuang, Nature (London) \textbf{%
402}, 390 (1999). A. S\o rensen and K. M\o lmer, Phys. Rev. A \textbf{58},
2745 (1998). X. Zhou, D. W. Leung, and I. L. Chuang, Phys. Rev. A \textbf{62}%
, 052316 (2000).

\bibitem{Eisert00} J. Eisert, \textit{et al.}, Phys. Rev. A \textbf{62},
052317 (2000).

\bibitem{Benjamin06} S. C. Benjamin, \textit{et al.}, N. J. Phys. \textbf{8}%
, 141 (2006).

\bibitem{Raussendorf00} R. Raussendorf and H. J. Briegel, Phys. Rev. Lett.
\textbf{86}, 5188 (2001).

\bibitem{Hume07} D. B. Hume, T. Rosenband and D. J. Wineland,
Phys. Rev. Lett. \textbf{99}, 120502 (2007).

\bibitem{JTSL07b} L. Jiang, \textit{et al.}, arXiv: \textbf{0709.4539}
(2007).

\bibitem{Meyn93} S. P. Meyn and R. L. Tweedie, \textit{Markov Chains and
Stochastic Stability} (Springer-Verlag, New York, 1993).

\bibitem{Garcia03} J. J. Garcia-Ripoll, P. Zoller, and J. I. Cirac, Phys.
Rev. Lett. 91, 157901 (2003).

\bibitem{Keller04} M. Keller, \textit{et al.}, Nature (London) \textbf{431},
1075 (2004).

\bibitem{Steinmetz06} T. Steinmetz, \textit{et al.}, App. Phys. Lett.
\textbf{89}, 111110 (2006).

\bibitem{Langer05} C. Langer, \textit{et al.}, Phys. Rev. Lett. \textbf{95},
060502 (2005).

\bibitem{Haffner05b} H. Haffner, \textit{et al.}, Appl. Phys. B \textbf{81},
151 (2005).

\bibitem{Knill05} E. Knill, Nature (London) \textbf{434}, 39 (2005).

\bibitem{Raussendorf07} R. Raussendorf and J. Harrington, Phys. Rev. Lett.
\textbf{98}, 190504 (2007).

\bibitem{Steane03} A. M. Steane, Phys. Rev. A 68, 042322 (2003).

\bibitem{Jiang07} L. Jiang, \textit{et al.}, Proc. Natl. Acad. Sci. U. S. A. \textbf{104}, 17291 (2007).
\end{thebibliography}
\end{document}